\documentclass[aps,prb,twocolumn,groupedaddress]{revtex4}
\usepackage{graphicx,latexsym}

\begin{document}

\title{Transverse depinning in weakly-pinned vortices}

\author{J.\ Lefebvre, M.\ Hilke, and Z.\ Altounian}

\affiliation{ Dpt.\ of Physics, McGill University, Montr\'eal, Canada
H3A 2T8.}

\begin{abstract}
We present experiments on weakly-pinned vortices, which exhibit 
a large critical transverse depinning force. These results are obtained in
the superconducting metallic glasses Fe$_{x}$Ni$_{1-x}$Zr$_{2}$ using crossed ac and dc driving currents.
We study the vortex depinning force due to the transverse ac drive as a
function of a longitudinal dc drive; the ac/dc combination permits
the separation of the transverse drive from the longitudinal one. We show
that the force required for depinning in the transverse direction is greatly
enhanced by the longitudinal drive, which demonstrates the existence of a
large transverse critical force. The measurements are performed as a function of magnetic field and temperature and show that the transverse critical force exists in a large portion of the phase diagram. Hysteresis observed at the transverse depinning threshold is consistent with a first-order transverse depinning transition.
\end{abstract}

\maketitle
\section{Introduction}
The vortex state of type II superconductors is rich in interaction phenomena: Governed by vortex-vortex repulsion, thermal fluctuations and pinning from material inhomogeneities, the competition between ordering and disordering gives rise to a wealth of static and dynamic phase transitions, as well as non-equilibrium phenomena. While the effect of disorder
on the static case has been widely studied in the past years
\cite{NattermannPRL64, BouchaudPRB46, BlatterRMP66, GiamarchiPRL72,
GiamarchiPRB52}, the driven case still has much to reveal; complications due to the need to consider many sources of anisotropy and non-linear elasticity, for instance, have impeded elucidation of the dynamic regime.

A large number of studies have demonstrated that at high driving forces, a
disordered system will show ordering due to averaging of disorder in the direction of motion: dynamical
ordering \cite{BhattacharyaPRL70, YaronPRL73, HellerqvistPRL76,
MarchevskyPRL78, ShiPRL67, MoonPRL77, RyuPRL77, OlsonPRL81, SpencerPRB55,
KoshelevPRL73, GiamarchiPRL76, BalentsPRL78, BalentsPRB57, LeDoussalPRB57,
PardoNature396, FangohrPRB63, FangohrPRB64, MarleyPRL74}. Experimentally, the crossover to
a more ordered vortex phase at large driving current is deduced in transport
measurements from the presence of a peak in the differential
resistance \cite{BhattacharyaPRL70, HellerqvistPRL76}. Dynamical ordering has also been revealed by a decrease of
the low frequency broadband noise \cite{MarleyPRL74} and an increase of the
longitudinal correlation length in neutron diffraction experiments\cite{YaronPRL73}; the phenomena has also been directly observed in magnetic decoration experiments\cite{MarchevskyPRL78, PardoNature396}.
Numerically and analytically, the
establishment that vortices undergo such dynamical phase transitions and
ordering has lead to the prediction of the existence of static channels in
which the vortices flow. These channels may be decoupled, in which case the
vortex phase obtained is the moving transverse glass (MTG) and is characterized by
smectic order; or they may be coupled, in which case one has the more ordered moving Bragg glass phase (MBG). Theoretical and numerical results have shown that these channels act as strong barriers against
transverse depinning, preventing transverse motion of longitudinally moving vortices from which stems the existence of a finite transverse critical
force \cite{MoonPRL77, OlsonPRL81, GiamarchiPRL76, LeDoussalPRB57, OlsonPRB61, RyuPRL77, FangohrPRB63, ReichhardtPRB65}. In addition to vortex systems in superconductors, other elastic media have been shown to exhibit a finite transverse depinning threshold: Simulations of elastic strings have revealed a hysteretic transverse depinning transition\cite{ReichhardtPRB65}, while experimental measurements of charge-density waves\cite{Markovic} and Wigner solids have evidenced transverse pinning\cite{Perruchot}. In particular, a transverse critical force of about one tenth the magnitude of the parallel critical force has been observed in a magnetically-induced Wigner solid in a GaAs/GaAlAs heterostructure\cite{Perruchot}. The existence of a transverse critical force in vortex lattices has not yet been observed experimentally, which is the focus of this article.

Here we present the results of an experimental study of the transverse depinning transition of vortices and
demonstrate the existence of a large transverse critical force and map out its phase diagram in field and drive. We
find that for a system driven longitudinally with a dc current,
application of a small transverse force, provided by an ac current,
does not result in immediate transverse depinning. In some regimes,
the transverse force required for depinning the vortices in the
transverse direction is even increased by more than 30 \% with
respect to the force required in the longitudinal case, thus
implying the appearance of very strong barriers against transverse
motion due to the longitudinal drive. This surprisingly large value can be compared to results from numerical studies which have found the ratio of the critical
transverse force to the critical longitudinal force
$f_{\perp}^{c}/f_{\parallel}^{c}$ to be of the order of 1 \%
\cite{MoonPRL77, RyuPRL77, OlsonPRB61} or 10 \% \cite{FangohrPRB63}.
Following findings from Fangohr \textit{et al.}\cite{FangohrPRB63}, this ratio is expected to
increase for more weakly-pinned vortices. However, finite size effects in
numerical simulations make very difficult and computationally expensive studies in the limit of very
weak pinning, which is the regime of our experiments. Indeed, in our
experiments, vortex pinning strength is at least six times smaller than has been used in Fangohr \textit{et al.}\cite{FangohrPRB63} and we
obtain a ratio $f_{\perp}^{c}/f_{\parallel}^{c}$, which can even exceed
100\%.

\section{Experimental techniques}
The measurements were performed on different samples of the metallic
glasses Fe$_{x}$Ni$_{1-x}$Zr$_{2}$ prepared by melt-spinning
\cite{AltounianPRB49} and which become superconducting below about
$2.4~$K depending on the iron content, and are
particularly clean with a very low critical current density $J_{c}\leq0.4~A/cm^{2}$ such that vortices are very weakly-pinned. The amorphous nature of the samples further guaranties the absence of long range order and isotropic pinning. These materials were found to be strong type II low temperature superconductors
\cite{HilkePRL91} from estimates of the different characterizing
length scales using standard expressions for superconductors in the
dirty limit \cite{KesPRB28}. They were also found to show a variety of phases
of longitudinal and transverse vortex motion, including a MBG-like
phase \cite{HilkePRL91, LefebvrePRB74}, and hence are ideal for the
study of transverse depinning.

\begin{figure}
[ptb]
\begin{center}
\includegraphics[
trim=0.233814in 0.432229in 0.201680in 0.000000in,
natheight=4.791900in,
natwidth=4.228100in,
height=3.7628in,
width=3.2785in]
{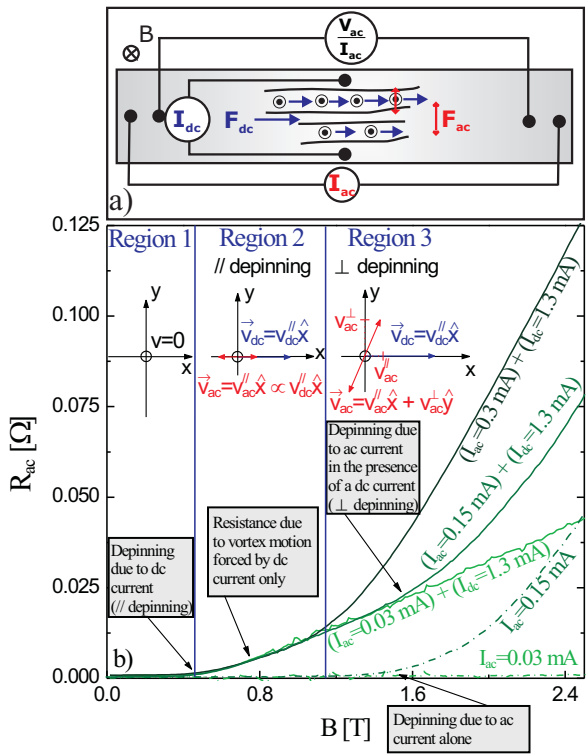}%
\caption{a) Drawing showing the contact configuration and resulting directions
of vortex motion. b) Resistance vs magnetic field measured with different
$I_{ac}$ and $I_{dc}$. The drawings show the trajectory followed by vortices
in the three regimes of vortex motion.}%
\label{RvsB}%
\end{center}
\end{figure}

We proceed by cooling the samples in a $^{3}$He system to a temperature below
0.4 K.  We use a dc current as the longitudinal drive, and a 15.9 Hz ac
current provided by a resistance bridge as the transverse drive. A magnetic field provided by a superconducting magnet and directed perpendicularly to the sample plane is also used. The
resistance is measured in the transverse direction with the resistance bridge.  Indium contacts are soldered to the sample in the configuration
shown in Fig.(\ref{RvsB}a).  In a magnetic
field perpendicular to the sample plane, the force exerted on the vortices by
the dc current $I_{dc}$ applied along the short edge of the sample acts in the
direction $\vec{F}_{dc}=\vec{J}_{dc}\times\vec{\Phi}$, such that the vortices
move under the action of this force along the long edge of the sample, as depicted. Similarly, the ac current $I_{ac}$ applied along the long edge of the sample induces a force which results in an oscillatory movement of the vortices in the direction parallel to the short edge of the sample. In this manner, the channels of
vortices are set up by the dc current in the longitudinal direction
along the long edge of the sample, and the transverse force is
provided by the ac current and directed along the short edge of the
sample.
 Evidently, the two sets of contacts used for dc and ac driving cannot be made perfectly perpendicular to each other, and the transverse
voltage measured (from the ac driving) also contains a component resulting from the ac
component along the dc longitudinally driven motion. This small contact
misalignment angle $\alpha$ ($\alpha$=0 if they are orthogonal) is sample dependent and can easily be taken into account as discussed below.

\section{Results}
\subsection{Identification of the purely transverse contribution}
Fig.\ref{RvsB}b) shows the
transverse ac resistance as a function of magnetic field for zero (dash-dotted lines)
and non-zero (solid lines) longitudinal dc currents. The dc driving current used is I$_{dc}$=1.3 mA. Three distinct regions are defined, corresponding to three different regimes of vortex motion: Region 1 is characterized by vortices pinned in both directions, as no matter the combination of ac and dc current used, none of the ac or dc current is strong enough to depin the vortices, leading
to zero resistance. In region 2, the resistance measured using solely ac currents I$_{ac}$=0.03 mA and I$_{ac}$=0.15 mA remains zero because these currents are not large enough to depin the vortices. However, the data acquired using a longitudinal dc current in addition to these transverse ac currents shows an ac resistance, which does not depend on the ac drives. Hence, this ac resistance results from the small component of vortex motion proportional to $\sin (\alpha)$ along the longitudinal (dc) direction, since depinning is associated with strong non-linearities of the V-I characteristics\cite{FeigelmanPRL63, FisherPRL62, NattermannPRL64}. This is in stark contrast to the behavior in region 3, where the ac
transverse resistance depends on the transverse ac current and
shown in region 3, where the vortices also move in the
transverse direction. Vectors representing the direction of vortex motion in these 3 regimes resulting from the combination of ac and dc driving are shown in Fig.\ref{RvsB}b). One then easily identifies the transverse depinning transition as the point in field and ac current where the ac
resistance starts to depend on the ac current. Hence, for a given
longitudinal dc drive, the resistance associated to the pure transverse dynamics can be obtained
as $R_{ac}^\perp(I_{ac},I_{dc})=R_{ac}(I_{ac},I_{dc})-R^\epsilon_{ac}(I^\epsilon_{ac},I_{dc})$, where $I^\epsilon_{ac}$ is a very small transverse ac measurement current which alone does not depin the vortices.
\subsection{V-I measurements}

\begin{figure}
[ptb]
\begin{center}
\includegraphics[
trim=0.458842in 5.297669in 0.508927in 0.000000in,
natheight=10.601700in,
natwidth=8.078200in,
height=2.348in,
width=3.1393in]%
{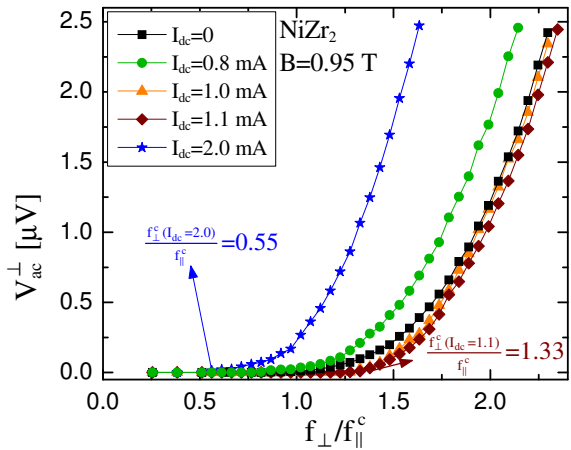}%
\caption{Transverse ac voltage versus applied ac force for different dc driven
cases normalized with the critical force in the static case. Error bars are smaller than the size of the dot. }%
\label{VvsFratio}%
\end{center}
\end{figure}

Transverse (ac) resistance measurements were performed on a sample of the alloy NiZr$_{2}$ at a fixed magnetic field B=0.95 T and temperature T=0.35 K for different fixed longitudinal (dc) driving currents. The measurements are performed by applying a dc current, to which a small ac current is added and increased in small steps, and the resulting ac voltage is recorded after each current step. The pure transverse voltages  are then obtained using the scheme described above, which removes the contribution due to the contact misalignment. In order to obtain the depinning forces we use an arbitrary cut-off voltage of $V_c=10$ $nV$ ($f^c\equiv f^c(V_c=10~nV)$), below which we consider the vortices as effectively pinned. The resulting V$^\perp_{ac}$ vs. $f_{\perp}/f_{\parallel}^{c}$ traces derived from these measurements are shown in Fig.\ref{VvsFratio}, where $f_{\perp}/f_{\parallel}^{c}$ is the transverse force normalized to the longitudinal depinning force. $f_{\parallel}^{c}$ is obtained from the transverse ac depinning current when the longitudinal current is set
to zero. This is valid since our system is isotropic and the transport properties measured using ac or dc currents are equivalent\cite{LefebvrePRB74}. The longitudinal depinning curve is thus represented by the I$_{dc}$=0 black curve in Fig.\ref{VvsFratio}, where the longitudinal depinning current was determined to be $I_c^{\parallel}$=0.55 mA. Because the contact misalignment is small we have $f_{\perp}/f_{\parallel}^{c}\simeq I_{ac}/I_c^{\parallel}$. 

In the presence of a longitudinal drive the results in Fig.\ref{VvsFratio} show that for I$_{dc}$=2 mA the transverse depinning force ($f_{\perp}^{c}$) is reduced as
compared to the longitudinal one, i.e. $f_{\perp}^{c}/f_{\parallel}^{c}<1$ as extracted from the corresponding intersect of V$^\perp_{ac}$ with 10 nV. In contrast, for 1.0 mA$\leq$I$_{dc}\leq$1.1 mA, the transverse depinning force is strongly enhanced by the presence of the longitudinal drive, with $f_{\perp}^{c}/f_{\parallel}^{c}$ reaching
$1.33$ for I$_{dc}$=1.1 mA. Hence, in this range of longitudinal drives, strong barriers against transverse motion are created by the longitudinal drive.

\subsection{The nature of criticality\label{criticality}}

\begin{figure}
[ptb]
\begin{center}
\includegraphics[
trim=0.731077in 5.133343in 0.670491in 0.000000in,
natheight=10.601700in,
natwidth=8.078200in,
height=2.4855in,
width=3.0303in]%
{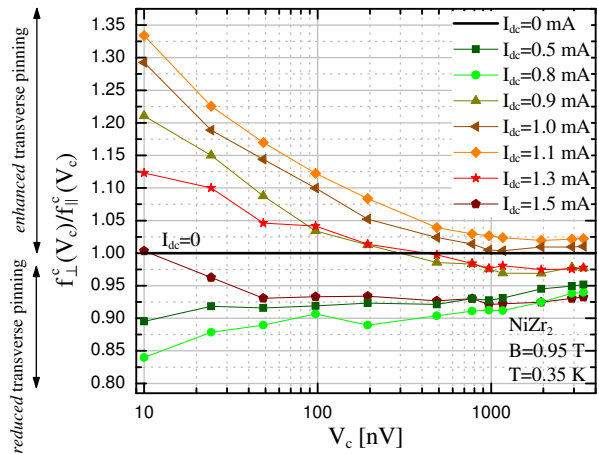}%
\caption{Ratio of the critical force in the driven case to that in the static
case as a function of transverse threshold voltage for different dc currents at
$B=0.95$ T. Error bars are smaller than the size of the dots.}%
\label{fratiovsvac}%
\end{center}
\end{figure}

Experimentally, defining a true critical depinning force is difficult, since vortex motion is detected by the produced voltage, which is noise limited. Moreover, non-zero temperatures allow for activated depinning, which smoothes out the depinning transition. However, here we compare two depinning mechanisms (transverse and longitudinal depinning) and by applying the same criteria on both their ratio should not depend on the voltage cutoff used as long as their dependence is similar. Quite strikingly, we find that the choice of cutoff voltage criteria greatly influences the value of $f_{\perp}^{c}(V_c)/f_{\parallel}^{c}(V_c)$ obtained. This is depicted in Fig.\ref{fratiovsvac}, which is obtained by changing the cutoff criteria. For all longitudinal driving currents, one observes that $f_{\perp}^{c}(V_c)/f_{\parallel}^{c}(V_c)$ approaches unity for large voltage cutoff. In fact, we obtain this ratio to be very close to one when the transverse force equals approximately five times
the longitudinal force (data not shown here). This implies that the barriers against transverse vortex motion
not only delay transverse depinning, but also constrain transverse
vortex motion at larger velocities as well, when the vortices are already depinned in the transverse direction. This effect was also
observed numerically by Fangohr \textit{et al.}\cite{FangohrPRB63}. More importantly, the data shown in
this figure can also be used to characterize the nature of the transverse depinning transition. Indeed, an increase of the ratio $f_{\perp}^{c}(V_c)/f_{\parallel}^{c}(V_c)$ with decreasing $V_c$ implies that the transverse pinning is more critical than the longitudinal one since absolute pinning turns on faster. We distinguish both behaviors as transverse enhanced pinning (when $\frac{d}{dV}f_{\perp}^{c}(V_c)/f_{\parallel}^{c}(V_c)|_{V_c=10nV}<0$) and as transverse reduced pinning otherwise. For curves with 0.9 mA $\leq I_{dc}\leq$1.5 mA, we observe strong transverse enhanced pinning in contrast to  $I_{dc}=0.5$ and 0.8 mA, where we observe transverse reduced pinning. The critical depinning forces are now defined as the depinning currents corresponding to the lowest voltage cutoff, i.e., $V_c=10$ nV.

\begin{figure}
[ptb]
\begin{center}
\includegraphics[
trim=0.000000in 4.989160in 1.016238in 0.000000in,
height=2.6351in,
width=3.3096in]%
{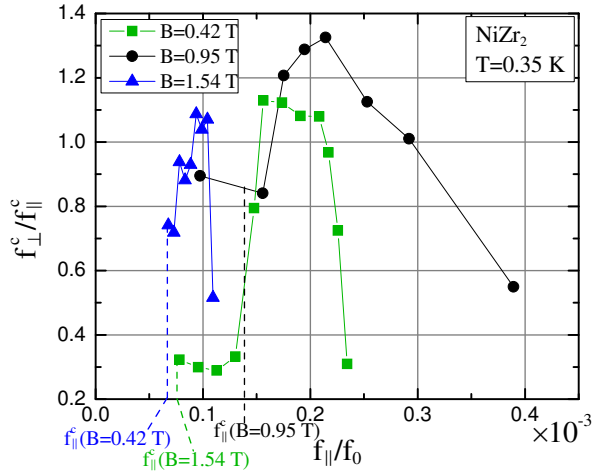}%
\caption{Ratio of the critical forces in the driven and the static case versus
the longitudinal dc force for different applied magnetic fields. Error bars are smaller than the size of the dots.}%
\label{fratiovsfdc}%
\end{center}
\end{figure}

This allows us to analyze the dependence of $f_{\perp}^c/f_{\parallel}^{c}$ on the longitudinal force for different vortex densities in Fig.\ref{fratiovsfdc}. Here the longitudinal force is presented in units of $f_{0}$, the interaction force between two vortices separated by a distance $\lambda$; in this manner, quantitative comparison of these results with those from numerical studies will be eased. The initial slight decrease of $f_{\perp}^{c}/f_{\parallel}^{c}$ in the region below $f_{\parallel}=0.16\times10^{-3}f_{0}$ (equivalent to I$_{dc}$=0.8 mA) for the B=0.95 T curve corresponds to a regime in which the longitudinal applied force is either smaller than the longitudinal depinning force, or just strong enough for depinning. The ratio $f_{\perp}^{c}/f_{\parallel}^{c}$ is thus dominated by the transverse motion and should therefore be close to one. However, the observed small initial decrease of
$f_{\perp}^{c}/f_{\parallel}^{c}$ is attributable to the longitudinal dc component in the transverse direction due to our contact misalignment, which now contributes to depinning in the transverse direction. For larger longitudinal force, a peak in $f_{\perp}^{c}/f_{\parallel}^{c}$ is seen, with a maximum at $f_{\parallel}=0.2\times10^{-3}f_{0}$. In this peak region, the longitudinal force is large enough to induce longitudinal depinning and induce vortex motion in channels; as a result, an important enhancement of the transverse depinning force is witnessed. For $f_{\parallel}>$0.33$\times10^{-3}$f$_{0}$, or I$_{dc}\geq$1.3 mA, the ratio of the critical forces starts to decay;
the decay gets stronger for I$_{dc}\geq$1.5 mA.  This decay
is likely due to additional dynamic disorder, which could weaken the
barriers against transverse vortex motion \cite{FangohrPRB63}. We have observe a similar
behavior in all the samples of Fe$_{x}$Ni$_{1-x}$Zr$_{2}$ with x=0, 0.1, 0.3, and 0.5, which hints that it is universal to this class of weakly-pinned superconductors.

\subsection{Comparison with numerical results}
A striking result obtained in this study is the huge magnitude of
$f_{\perp}^{c}/f_{\parallel}^{c}$, ranging from 0.55 to 1.33, compared to between 0.01 and 0.1 obtained in numerical studies \cite{MoonPRL77, RyuPRL77,
OlsonPRB61, FangohrPRB63}. This much larger
magnitude of the normalized critical transverse depinning force in
our experimental weakly pinned system is consistent with results from a numerical study by Fangohr \textit{et al.} \cite{FangohrPRB63}, in which a larger critical forces ratio is obtained for more weakly-pinned
samples. This result was attributed to the independence of the transverse depinning on the longitudinal pinning potential. As a result, for weak longitudinal pinning, $f_{\parallel}^{c}$ becomes very small, but $f_{\perp}^{c}$ remains essentially unaffected by the degree of longitudinal pinning since the moving system is topologically ordered.
To compare quantitatively our experimental results with those from numerical studies, we obtain the pinning force per unit length for our sample from $f_{p}=A\left\vert \vec{J}_{c}\times\vec{B}\right\vert $, where $A$ is the area of the sample
perpendicular to the $B$ field. This leads to $f_{p}=0.02f_{0}$, which means
it is 6 times less pinned than the most weakly-pinned sample simulated
in Ref.\cite{FangohrPRB63}.  In addition, we obtain the critical
longitudinal depinning force for our system using the V=10 nV
cutoff to be $f_{\parallel}^{c}=1\times10^{-4}f_{0}$, which is more than 200
times smaller than the longitudinal depinning force simulated in
Ref.\cite{FangohrPRB63}. These quantities confirm the weak-pinning
nature of our samples, and explain the very large observed
critical transverse to longitudinal force ratio.\\
Another particularity of our results compared to results from numerical
studies\cite{GiamarchiPRL76, LeDoussalPRB57,
FangohrPRB63, ReichhardtPRB65} is the increase of the transverse critical force with increasing
longitudinal drive in the critical region, for instance between
$f_{\parallel}=0.16\times10^{-3}f_{0}$ and  $f_{\parallel}=0.22\times10^{-3}f_{0}$ in Fig.\ref{fratiovsfdc}a). This result is contrary to the observation in
numerical simulations of a strong decrease of the transverse critical force
with increasing longitudinal vortex velocity \cite{GiamarchiPRL76, LeDoussalPRB57,
FangohrPRB63, ReichhardtPRB65} attributable to the fact that for large longitudinal vortex velocity, the channels become straighter and cannot go through optimal pinning sites as easily, with result that the transverse barriers are weakened. We speculate that we obtain our differing results because pinning in these samples is dominated by collective effects rather than by the underlying pinning potential.

\subsection{Magnetic phase diagram}

\begin{figure}
[tbh]
\begin{center}
\includegraphics[
trim=0.488731in 5.766264in 0.980694in 0.000000in,
natheight=10.601700in,
natwidth=8.078200in,
height=2.13in,
width=2.9in]%
{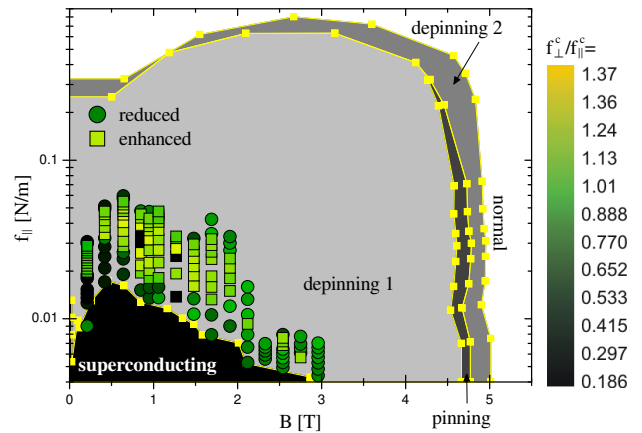}%
\caption{Longitudinal driving force vs magnetic field phase diagram showing
the different vortex phases observed in the weakly-pinned amorphous metallic
glass NiZr$_{2}$ and determined from longitudinal resistance measurements
performed using an ac driving current. Also shown are regions for which the transverse depinning force was determined to have magnitude $f_{\perp}^{c}/f_{\parallel}^{c}$ given according to the color scale on the
right. Squares represent regions for which transverse depinning is enhanced with respect to longitudinal depinning, whereas circles represent regions where it is found to be reduced.}%
\label{Bphase}%
\end{center}
\end{figure}

We further investigated the transverse depinning transition for different
applied magnetic fields; since increasing the magnetic field is equivalent to
increasing effective disorder and thus pinning, this basically amounts to studying
transverse depinning as a function of longitudinal pinning properties. The
results reveal that a large critical transverse depinning force exists in
regions of the phase diagram where a large longitudinal pinning force also
exists. This is emphasized in Fig.\ref{Bphase} which shows the complete
longitudinal driving force $f_{\parallel}-$magnetic field phase diagram extracted from resistance measurements as a function of magnetic field. Designation of the different phases is as described by Hilke \textit{et al.}\cite{HilkePRL91}; depinning 1 and 2 correspond to depinned vortex states
reminiscent of the MBG and the MTG \cite{LefebvrePRB74} respectively, while
the pinning phase corresponds to the peak effect \cite{HilkePRL91}. The graph
also displays data points showing regions where transverse depinning was
investigated and found to be enhanced (circles) with the magnitude of
$f_{\perp}^{c}/f_{\parallel}^{c}$ represented by the color scale on the right), and reduced
 (squares). The longitudinal depinning force, represented in this
graph by the yellow line delimiting the superconducting and depinning 1
phases, is found to increase for low magnetic field ($B\leq0.5$ T) as disorder proliferates, but decreases for larger fields ($0.5\leq B\leq3$ T).

Since pinning in these samples is so weak and depends mostly on collective
effects rather than on the underlying disorder, increase of the magnetic field
passed about 0.5 T does not result in an increase of the pinning force due to the disorder it
generates; on the contrary, this regime witnesses an decrease of the
driving force necessary to depin the vortex lattice. It can also be seen that while the transverse depinning
transition could only be investigated in the region $0\leq B\leq3$ T because the depinning current becomes too low for larger fields and noise
dominates the signal, a large transverse depinning force, with $f_{\perp}%
^{c}/f_{\parallel}^{c}$ significantly larger than 1, is only found in the region
$0.4\leq B\leq1.6$ T where the longitudinal depinning force is also largest. This can be
understood from consideration of the fact that the larger the longitudinal pinning
force is (or equivalently the longitudinal pinning potential), the stronger
the channels must be; the stronger channels must then also constitute a
stronger barrier against transverse depinning and lead to the large transverse
critical force. Of course, it has to be kept in mind that this is only true in the elastic regime of vortex motion in which the MBG is well-developed; for larger fields, the MBG is weakened (but does not necessarily break down) due to the proliferation of defects and installment of plastic flow, then resulting in a reduced $f_{\perp}^{c}/f_{\parallel}^{c}$ with increasing $f_{\parallel}^{c}$. A similar increase of the transverse critical force with
increasing longitudinal pinning strength for a simulated vortex lattice in the
elastic flow regime has also been observed by Fangohr \textit{et al.}
\cite{FangohrPRB63}.

It is important to note that the transverse depinning force data presented in
Fig.\ref{Bphase} does not necessarily represent the boundary of the critical
transverse depinning phase; for instance, for $B\simeq1$ T, no longitudinal force larger than $f_{\parallel}=0.048$ N/m was investigated but the transverse depinning transition is still found to
be critical for this force such that the critical region might extend to
larger longitudinal force at this field. However, for $B=2.96$ T which is the largest field probed, no region showing enhanced transverse
depinning was found.

\subsection{Temperature phase diagram}

\begin{figure}
[tbh]
\begin{center}
\includegraphics[
trim=0.000000in 0.488738in 0.491155in 0.000000in,
natheight=10.601700in,
natwidth=8.078200in,
height=4.1736in,
width=3.1393in]%
{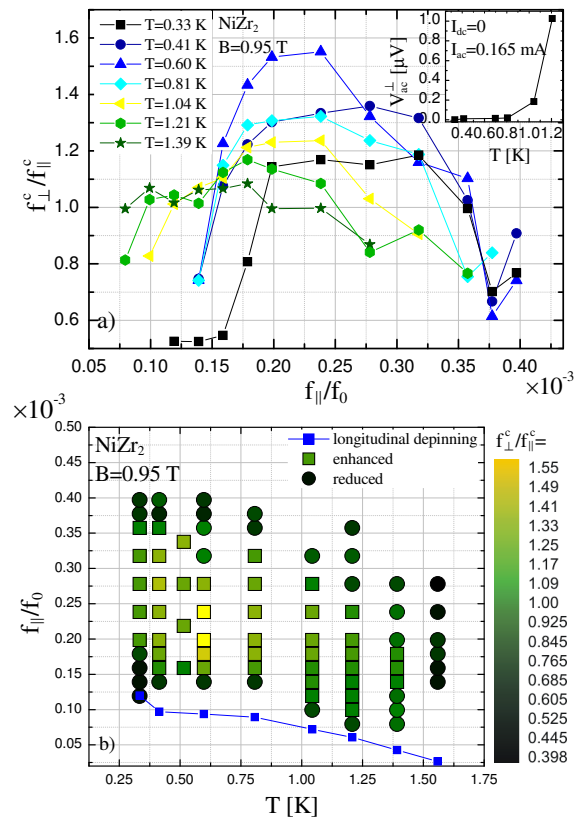}%
\caption{a) Ratio of the critical transverse current determined with a 10 nV criterion in the driven and the static case vs the
longitudinal driving current for a sample of NiZr$_{2}$ at B=0.95 T and for different temperatures. Inset: Transverse
dissipation voltage as a function of temperature for zero I$_{dc}=0$ and
I$_{ac}=0.165$ mA. The data is extracted from measurements of
V$_{ac}$ as a function of I$_{ac}$ at different T. b) Map of the critical
current ratio in the driven and static case according to the color scale, for
different longitudinal driving forces and magnetic fields. The squares
represent enhanced transverse pinning and the circles reduced transverse pinning with respect to longitudinal pinning. The solid line represents the longitudinal depinning line.}%
\label{Tphase}%
\end{center}
\end{figure}

With increasing temperature, the MBG phase has been predicted to survive up to
the vortex lattice melting temperature \cite{LeDoussalPRB57}, with
accompanying vortex motion in channels and critical transverse force
\cite{FangohrPRB63}; a broadening of the vortex channels due to thermal
displacements about their average position with a weakening of the transverse
barriers is however anticipated. Therefore, while longitudinal depinning is
normally eased at high temperature because vortices have more energy to
overcome the pinning potential, transverse depinning should also proceed more
readily because the vortices have more energy to surmount the channel
barriers, which should also be weakened because of the
thermally-induced lateral motion of the vortices. This is however not exactly
the scenario observed from investigation of the transverse depinning force for
different longitudinal driving currents as a function of temperature below
T$_{c}$. Indeed, as can be seen from the $f_{\perp}^{c}/f_{\parallel}^{c}$ vs longitudinal current $f_{\parallel}$ data presented in Fig.\ref{Tphase}a),
it is found that $f_{\perp}^{c}/f_{\parallel}^{c}$ reaches a larger maximal value for T=0.41 K than for T=0.33 K, and even attains its maximum at T=0.60 K with $f_{\perp}^{c}/f_{\parallel}^{c}$=1.55. This observation signifies that for identical applied longitudinal force,
barriers against transverse vortex motion are strengthened by the increase in
temperature, whereas they are seen to weaken in some numerical simulations \cite{FangohrPRB63}. This can be understood because the longitudinal vortex velocity increases with increasing temperature when induced by an equal external driving force (see inset of Fig.\ref{Tphase}a), recalling that voltage is proportional to vortex velocity), and hence results in a strengthening of the
barriers, similar to what is observed for increasing longitudinal drive.
Therefore, it appears from these results that for temperatures between 0.4 K and 0.6 K, increased transverse pinning due to increased longitudinal vortex velocity is stronger than the temperature-dependent weakening of the channel barriers.%

The results of the transverse critical current as a function of temperature
and longitudinal driving current are summarized in Fig.\ref{Tphase}b in
which the blue line represents the longitudinal depinning current determined
with a 10 nV criterion, and the circles and squares represent the magnitude of the transverse to
longitudinal critical current ratio according to the colors shown in the
legend. T$_{c}$ for this sample is 2.4 K, but the largest temperature investigated is 1.56 K, well below T$_{c}$; higher temperatures could not be investigated because
at such temperatures the depinning current becomes too low and the measured
signal becomes too noisy. Therefore, the region of the $f_{\parallel}$-T phase diagram
shown to exhibit critical transverse depinning is not bounded and a region
exhibiting critical transverse depinning probably might exist for higher
temperatures below T$_{c}$.

\subsection{History effects}

The experiment described above raises interesting issues about history
effects. For instance, as pointed out by Le Doussal and Giamarchi
\cite{LeDoussalPRB57}, one can wonder if the effect of applying a force in
perpendicular directions simultaneously (i.e. $F=f_{\parallel}\widehat{x}%
+f_{\perp}\widehat{y}$) is the same as first applying the longitudinal force
$f_{\parallel}\widehat{x}\,$, then waiting for steady state before applying the
transverse force $f_{\perp}\widehat{y}$, as we did in this experiment. As
discussed in Ref.\cite{LeDoussalPRB57}, if the vortex state were liquid-like,
the result of both these experiments would be identical, but for the moving
vortex glass, the answer is not trivial and could bear important information
about the glassy state. As an observation of history effects, Reichhardt and Olson \cite{ReichhardtPRB65} have obtained, in a numerical study of transverse dynamics of elastic
strings, hysteretic
transverse depinning transitions for strings also driven in the parallel
direction.

\begin{figure}
[tbh]
\begin{center}
\includegraphics[
trim=0.000000in 4.587356in 0.508927in 0.000000in,
natheight=10.601700in,
natwidth=8.078200in,
height=2.6584in,
width=3.3399in]%
{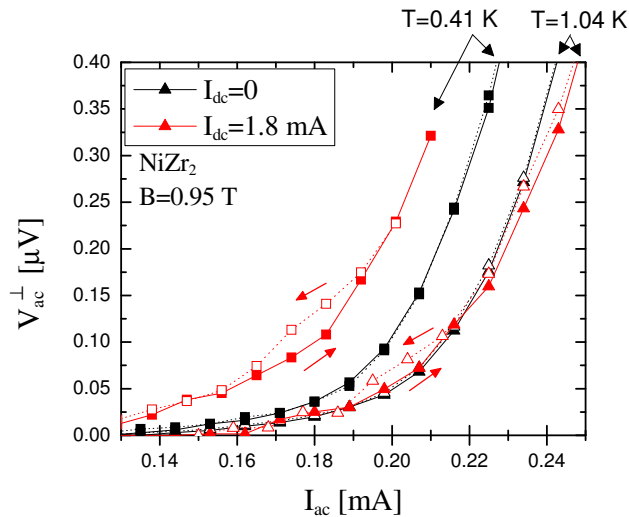}%
\caption{Transverse voltage as a function of transverse current at
B=0.95 T for different longitudinal driving currents at T=0.41 K and T=1.04 K. The solid lines with
filled data points were acquired by increasing I$_{ac}$, while the dotted lines
with empty data points are for decreasing I$_{ac}$.}%
\label{hysteresis}%
\end{center}
\end{figure}

With the aim of studying possible history effects experimentally in the
transverse depinning transition, we have performed some measurements of
V$_{ac}$ as a function of transverse driving current I$_{ac}$ in the presence
of a longitudinal driving current I$_{dc}$ for both increasing and decreasing
I$_{ac}$. The data was obtained in the following manner: First I$_{dc}$ is
applied, then when the measured voltage is stable and steady state is
presumably reached, a small I$_{ac}$ is also applied. I$_{ac}$ is then increased
in small steps and the resulting V$_{ac}$ is recorded. When V$_{ac}$ is large
and we are certain that the vortices are well depinned in the transverse
direction, we start decreasing I$_{ac}$, still in small steps, and recording
V$_{ac}$ when steady-state is established after each change of I$_{ac}$. Some
of the data acquired in this manner is shown in Fig.\ref{hysteresis} at
T=0.41 K and T=1.04 K for dc drives I$_{dc}=0$ and I$_{dc}=1.8$ mA. In the figure,
the solid lines represent data acquired by increasing I$_{ac}$ while the dotted
lines are for decreasing I$_{ac}$. A region showing hysteresis is observed
close to the depinning transition in cases also driven longitudinally with I$_{dc}=1.8$ mA. The size of the hysteresis loop also seems to increase with increasing
temperature. In opposition, the case I$_{dc}=0$, which represents longitudinal
depinning, is definitely free of hysteresis, as it should be. The
counterclockwise direction of the hysteresis loops in the V-I$_{ac}$
characteristics observed experimentally here is the same as that observed
numerically in Ref.\cite{ReichhardtPRB65}. This observation of hysteresis at
the transverse depinning transition is the first experimental confirmation
that this transition is consistent with a first order transition and not merely a crossover.%

\section{Summary}
We have investigated the transverse depinning transition in weakly-pinned
vortices in the superconducting metallic glass NiZr$_{2}$ from electrical transport measurements using ac and dc
currents in crossed configurations. We have obtained that for a vortex system
driven longitudinally, application of a small transverse force does not result
in immediate transverse motion, thus confirming experimentally for the first
time the existence of a critical transverse depinning force in such systems.
In some cases, the force required for depinning in the transverse direction is
increased by more than 30 \% with respect to the force required in the
longitudinal case, thus implying that barriers against transverse motion in the
MBG phase are very strong in these weakly-pinned vortex systems. \\
The transverse depinning transition was further investigated for various magnetic
fields and temperatures and was found to exhibit critical behavior in a large
portion of the phase diagram. From these results, it was deduced that a large
transverse critical force exists in regions of the magnetic phase diagram
where a large longitudinal depinning force also exists. It was also shown that
on the contrary to what is generally observed in simulations
\cite{LeDoussalPRB57, GiamarchiPRL76, FangohrPRB63, ReichhardtPRB65},
increasing longitudinal drive or longitudinal vortex velocity can result in an
increase of the transverse depinning force. We speculate that this effect, as
well as the large transverse depinning force obtained here compared to
$0.01\leq f_{\perp}^{c}/f_{\parallel}^{c}\leq0.1$ in simulations \cite{MoonPRL77,
RyuPRL77, FangohrPRB63, OlsonPRB61} are both attributable to the very
weak-pinning character of our metallic glasses. In the former case, the
weak-pinning properties infer that pinning is mainly governed by collective
effects and interaction between vortices than by intrinsic disorder, such that
the weakening of barriers and straightening of channels witnessed as the
channels fail to adapt to the pinning potential at large vortex velocity in
numerical simulations \cite{LeDoussalPRB57, GiamarchiPRL76, FangohrPRB63,
ReichhardtPRB65} does not take place. Similarly, the large $f_{\perp}^{c}%
/f_{\parallel}^{c}$ observed results from the fact that the transverse depinning force
is largely independent of longitudinal pinning \cite{FangohrPRB63}, such that
for a very weakly-pinned system the ratio $f_{\perp}^{c}/f_{\parallel}^{c}$ can be greatly
enhanced.  \\
Hysteresis close to the transverse depinning transition was also
observed, thus confirming experimentally  the first-order
nature of the transverse depinning transition in the glass state.

\bibliography{ACDCcurrentsJuly16}

\begin{thebibliography}{33}
\expandafter\ifx\csname natexlab\endcsname\relax\def\natexlab#1{#1}\fi
\expandafter\ifx\csname bibnamefont\endcsname\relax
  \def\bibnamefont#1{#1}\fi
\expandafter\ifx\csname bibfnamefont\endcsname\relax
  \def\bibfnamefont#1{#1}\fi
\expandafter\ifx\csname citenamefont\endcsname\relax
  \def\citenamefont#1{#1}\fi
\expandafter\ifx\csname url\endcsname\relax
  \def\url#1{\texttt{#1}}\fi
\expandafter\ifx\csname urlprefix\endcsname\relax\def\urlprefix{URL }\fi
\providecommand{\bibinfo}[2]{#2}
\providecommand{\eprint}[2][]{\url{#2}}

\bibitem[{\citenamefont{Nattermann}(1990)}]{NattermannPRL64}
\bibinfo{author}{\bibfnamefont{T.}~\bibnamefont{Nattermann}},
  \bibinfo{journal}{Phys. Rev. Lett.} \textbf{\bibinfo{volume}{64}},
  \bibinfo{pages}{2454} (\bibinfo{year}{1990}).

\bibitem[{\citenamefont{Bouchaud et~al.}(1992)\citenamefont{Bouchaud, M\'ezard,
  and Yedidia}}]{BouchaudPRB46}
\bibinfo{author}{\bibfnamefont{J.-P.} \bibnamefont{Bouchaud}},
  \bibinfo{author}{\bibfnamefont{M.}~\bibnamefont{M\'ezard}}, \bibnamefont{and}
  \bibinfo{author}{\bibfnamefont{J.~S.} \bibnamefont{Yedidia}},
  \bibinfo{journal}{Phys. Rev. B} \textbf{\bibinfo{volume}{46}},
  \bibinfo{pages}{14686} (\bibinfo{year}{1992}).

\bibitem[{\citenamefont{Blatter et~al.}(1994)\citenamefont{Blatter, Feigel'man,
  Geshkenbein, Larkin, and Vinokur}}]{BlatterRMP66}
\bibinfo{author}{\bibfnamefont{G.}~\bibnamefont{Blatter}},
  \bibinfo{author}{\bibfnamefont{M.~V.} \bibnamefont{Feigel'man}},
  \bibinfo{author}{\bibfnamefont{V.~B.} \bibnamefont{Geshkenbein}},
  \bibinfo{author}{\bibfnamefont{A.~I.} \bibnamefont{Larkin}},
  \bibnamefont{and} \bibinfo{author}{\bibfnamefont{V.~M.}
  \bibnamefont{Vinokur}}, \bibinfo{journal}{Rev. Mod. Phys.}
  \textbf{\bibinfo{volume}{66}}, \bibinfo{pages}{1125} (\bibinfo{year}{1994}).

\bibitem[{\citenamefont{Giamarchi and Le~Doussal}(1994)}]{GiamarchiPRL72}
\bibinfo{author}{\bibfnamefont{T.}~\bibnamefont{Giamarchi}} \bibnamefont{and}
  \bibinfo{author}{\bibfnamefont{P.}~\bibnamefont{Le~Doussal}},
  \bibinfo{journal}{Phys. Rev. Lett.} \textbf{\bibinfo{volume}{72}},
  \bibinfo{pages}{1530} (\bibinfo{year}{1994}).

\bibitem[{\citenamefont{Giamarchi and Le~Doussal}(1995)}]{GiamarchiPRB52}
\bibinfo{author}{\bibfnamefont{T.}~\bibnamefont{Giamarchi}} \bibnamefont{and}
  \bibinfo{author}{\bibfnamefont{P.}~\bibnamefont{Le~Doussal}},
  \bibinfo{journal}{Phys. Rev. B} \textbf{\bibinfo{volume}{52}},
  \bibinfo{pages}{1242} (\bibinfo{year}{1995}).

\bibitem[{\citenamefont{Bhattacharya and Higgins}(1993)}]{BhattacharyaPRL70}
\bibinfo{author}{\bibfnamefont{S.}~\bibnamefont{Bhattacharya}}
  \bibnamefont{and} \bibinfo{author}{\bibfnamefont{M.~J.}
  \bibnamefont{Higgins}}, \bibinfo{journal}{Phys. Rev. Lett.}
  \textbf{\bibinfo{volume}{70}}, \bibinfo{pages}{2617} (\bibinfo{year}{1993}).

\bibitem[{\citenamefont{Yaron et~al.}(1994)\citenamefont{Yaron, Gammel, Huse,
  Kleiman, Oglesby, Bucher, Batlogg, Bishop, Mortensen, Clausen
  et~al.}}]{YaronPRL73}
\bibinfo{author}{\bibfnamefont{U.}~\bibnamefont{Yaron}},
  \bibinfo{author}{\bibfnamefont{P.~L.} \bibnamefont{Gammel}},
  \bibinfo{author}{\bibfnamefont{D.~A.} \bibnamefont{Huse}},
  \bibinfo{author}{\bibfnamefont{R.~N.} \bibnamefont{Kleiman}},
  \bibinfo{author}{\bibfnamefont{C.~S.} \bibnamefont{Oglesby}},
  \bibinfo{author}{\bibfnamefont{E.}~\bibnamefont{Bucher}},
  \bibinfo{author}{\bibfnamefont{B.}~\bibnamefont{Batlogg}},
  \bibinfo{author}{\bibfnamefont{D.~J.} \bibnamefont{Bishop}},
  \bibinfo{author}{\bibfnamefont{K.}~\bibnamefont{Mortensen}},
  \bibinfo{author}{\bibfnamefont{K.}~\bibnamefont{Clausen}},
  \bibnamefont{et~al.}, \bibinfo{journal}{Phys. Rev. Lett.}
  \textbf{\bibinfo{volume}{73}}, \bibinfo{pages}{2748} (\bibinfo{year}{1994}).

\bibitem[{\citenamefont{Hellerqvist et~al.}(1996)\citenamefont{Hellerqvist,
  Ephron, White, Beasley, and Kapitulnik}}]{HellerqvistPRL76}
\bibinfo{author}{\bibfnamefont{M.~C.} \bibnamefont{Hellerqvist}},
  \bibinfo{author}{\bibfnamefont{D.}~\bibnamefont{Ephron}},
  \bibinfo{author}{\bibfnamefont{W.~R.} \bibnamefont{White}},
  \bibinfo{author}{\bibfnamefont{M.~R.} \bibnamefont{Beasley}},
  \bibnamefont{and}
  \bibinfo{author}{\bibfnamefont{A.}~\bibnamefont{Kapitulnik}},
  \bibinfo{journal}{Phys. Rev. Lett.} \textbf{\bibinfo{volume}{76}},
  \bibinfo{pages}{4022} (\bibinfo{year}{1996}).

\bibitem[{\citenamefont{Marchevsky et~al.}(1997)\citenamefont{Marchevsky,
  Aarts, Kes, and Indenbom}}]{MarchevskyPRL78}
\bibinfo{author}{\bibfnamefont{M.}~\bibnamefont{Marchevsky}},
  \bibinfo{author}{\bibfnamefont{J.}~\bibnamefont{Aarts}},
  \bibinfo{author}{\bibfnamefont{P.~H.} \bibnamefont{Kes}}, \bibnamefont{and}
  \bibinfo{author}{\bibfnamefont{M.~V.} \bibnamefont{Indenbom}},
  \bibinfo{journal}{Phys. Rev. Lett.} \textbf{\bibinfo{volume}{78}},
  \bibinfo{pages}{531} (\bibinfo{year}{1997}).

\bibitem[{\citenamefont{Shi and Berlinsky}(1991)}]{ShiPRL67}
\bibinfo{author}{\bibfnamefont{A.-C.} \bibnamefont{Shi}} \bibnamefont{and}
  \bibinfo{author}{\bibfnamefont{A.~J.} \bibnamefont{Berlinsky}},
  \bibinfo{journal}{Phys. Rev. Lett.} \textbf{\bibinfo{volume}{67}},
  \bibinfo{pages}{1926} (\bibinfo{year}{1991}).

\bibitem[{\citenamefont{Moon et~al.}(1996)\citenamefont{Moon, Scalettar, and
  Zim\'anyi}}]{MoonPRL77}
\bibinfo{author}{\bibfnamefont{K.}~\bibnamefont{Moon}},
  \bibinfo{author}{\bibfnamefont{R.~T.} \bibnamefont{Scalettar}},
  \bibnamefont{and} \bibinfo{author}{\bibfnamefont{G.~T.}
  \bibnamefont{Zim\'anyi}}, \bibinfo{journal}{Phys. Rev. Lett.}
  \textbf{\bibinfo{volume}{77}}, \bibinfo{pages}{2778} (\bibinfo{year}{1996}).

\bibitem[{\citenamefont{Ryu et~al.}(1996)\citenamefont{Ryu, Hellerqvist,
  Doniach, Kapitulnik, and Stroud}}]{RyuPRL77}
\bibinfo{author}{\bibfnamefont{S.}~\bibnamefont{Ryu}},
  \bibinfo{author}{\bibfnamefont{M.}~\bibnamefont{Hellerqvist}},
  \bibinfo{author}{\bibfnamefont{S.}~\bibnamefont{Doniach}},
  \bibinfo{author}{\bibfnamefont{A.}~\bibnamefont{Kapitulnik}},
  \bibnamefont{and} \bibinfo{author}{\bibfnamefont{D.}~\bibnamefont{Stroud}},
  \bibinfo{journal}{Phys. Rev. Lett.} \textbf{\bibinfo{volume}{77}},
  \bibinfo{pages}{5114} (\bibinfo{year}{1996}).

\bibitem[{\citenamefont{Olson et~al.}(1998)\citenamefont{Olson, Reichhardt, and
  Nori}}]{OlsonPRL81}
\bibinfo{author}{\bibfnamefont{C.~J.} \bibnamefont{Olson}},
  \bibinfo{author}{\bibfnamefont{C.}~\bibnamefont{Reichhardt}},
  \bibnamefont{and} \bibinfo{author}{\bibfnamefont{F.}~\bibnamefont{Nori}},
  \bibinfo{journal}{Phys. Rev. Lett.} \textbf{\bibinfo{volume}{81}},
  \bibinfo{pages}{3757} (\bibinfo{year}{1998}).

\bibitem[{\citenamefont{Spencer and Jensen}(1997)}]{SpencerPRB55}
\bibinfo{author}{\bibfnamefont{S.}~\bibnamefont{Spencer}} \bibnamefont{and}
  \bibinfo{author}{\bibfnamefont{H.~J.} \bibnamefont{Jensen}},
  \bibinfo{journal}{Phys. Rev. B} \textbf{\bibinfo{volume}{55}},
  \bibinfo{pages}{8473} (\bibinfo{year}{1997}).

\bibitem[{\citenamefont{Koshelev and Vinokur}(1994)}]{KoshelevPRL73}
\bibinfo{author}{\bibfnamefont{A.~E.} \bibnamefont{Koshelev}} \bibnamefont{and}
  \bibinfo{author}{\bibfnamefont{V.~M.} \bibnamefont{Vinokur}},
  \bibinfo{journal}{Phys. Rev. Lett.} \textbf{\bibinfo{volume}{73}},
  \bibinfo{pages}{3580} (\bibinfo{year}{1994}).

\bibitem[{\citenamefont{Giamarchi and Le~Doussal}(1996)}]{GiamarchiPRL76}
\bibinfo{author}{\bibfnamefont{T.}~\bibnamefont{Giamarchi}} \bibnamefont{and}
  \bibinfo{author}{\bibfnamefont{P.}~\bibnamefont{Le~Doussal}},
  \bibinfo{journal}{Phys. Rev. Lett.} \textbf{\bibinfo{volume}{76}},
  \bibinfo{pages}{3408} (\bibinfo{year}{1996}).

\bibitem[{\citenamefont{Balents et~al.}(1997)\citenamefont{Balents, Marchetti,
  and Radzihovsky}}]{BalentsPRL78}
\bibinfo{author}{\bibfnamefont{L.}~\bibnamefont{Balents}},
  \bibinfo{author}{\bibfnamefont{M.~C.} \bibnamefont{Marchetti}},
  \bibnamefont{and}
  \bibinfo{author}{\bibfnamefont{L.}~\bibnamefont{Radzihovsky}},
  \bibinfo{journal}{Phys. Rev. Lett.} \textbf{\bibinfo{volume}{78}},
  \bibinfo{pages}{751} (\bibinfo{year}{1997}).

\bibitem[{\citenamefont{Balents et~al.}(1998)\citenamefont{Balents, Marchetti,
  and Radzihovsky}}]{BalentsPRB57}
\bibinfo{author}{\bibfnamefont{L.}~\bibnamefont{Balents}},
  \bibinfo{author}{\bibfnamefont{M.~C.} \bibnamefont{Marchetti}},
  \bibnamefont{and}
  \bibinfo{author}{\bibfnamefont{L.}~\bibnamefont{Radzihovsky}},
  \bibinfo{journal}{Phys. Rev. B} \textbf{\bibinfo{volume}{57}},
  \bibinfo{pages}{7705} (\bibinfo{year}{1998}).

\bibitem[{\citenamefont{Le~Doussal and Giamarchi}(1998)}]{LeDoussalPRB57}
\bibinfo{author}{\bibfnamefont{P.}~\bibnamefont{Le~Doussal}} \bibnamefont{and}
  \bibinfo{author}{\bibfnamefont{T.}~\bibnamefont{Giamarchi}},
  \bibinfo{journal}{Phys. Rev. B} \textbf{\bibinfo{volume}{57}},
  \bibinfo{pages}{11356} (\bibinfo{year}{1998}).

\bibitem[{\citenamefont{Pardo et~al.}(1998)\citenamefont{Pardo, De~La~Cruz,
  Gammel, Bucher, and Bishop}}]{PardoNature396}
\bibinfo{author}{\bibfnamefont{F.}~\bibnamefont{Pardo}},
  \bibinfo{author}{\bibfnamefont{F.}~\bibnamefont{De~La~Cruz}},
  \bibinfo{author}{\bibfnamefont{P.~L.} \bibnamefont{Gammel}},
  \bibinfo{author}{\bibfnamefont{E.}~\bibnamefont{Bucher}}, \bibnamefont{and}
  \bibinfo{author}{\bibfnamefont{D.~J.} \bibnamefont{Bishop}},
  \bibinfo{journal}{Nature} \textbf{\bibinfo{volume}{396}},
  \bibinfo{pages}{348} (\bibinfo{year}{1998}).

\bibitem[{\citenamefont{Fangohr
  et~al.}(2001{\natexlab{a}})\citenamefont{Fangohr, de~Groot, and
  Cox}}]{FangohrPRB63}
\bibinfo{author}{\bibfnamefont{H.}~\bibnamefont{Fangohr}},
  \bibinfo{author}{\bibfnamefont{P.~A.~J.} \bibnamefont{de~Groot}},
  \bibnamefont{and} \bibinfo{author}{\bibfnamefont{S.~J.} \bibnamefont{Cox}},
  \bibinfo{journal}{Phys. Rev. B} \textbf{\bibinfo{volume}{63}},
  \bibinfo{pages}{064501} (\bibinfo{year}{2001}{\natexlab{a}}).

\bibitem[{\citenamefont{Fangohr
  et~al.}(2001{\natexlab{b}})\citenamefont{Fangohr, Cox, and
  de~Groot}}]{FangohrPRB64}
\bibinfo{author}{\bibfnamefont{H.}~\bibnamefont{Fangohr}},
  \bibinfo{author}{\bibfnamefont{S.~J.} \bibnamefont{Cox}}, \bibnamefont{and}
  \bibinfo{author}{\bibfnamefont{P.~A.~J.} \bibnamefont{de~Groot}},
  \bibinfo{journal}{Phys. Rev. B} \textbf{\bibinfo{volume}{64}},
  \bibinfo{pages}{064505} (\bibinfo{year}{2001}{\natexlab{b}}).

\bibitem[{\citenamefont{Marley et~al.}(1995)\citenamefont{Marley, Higgins, and
  Bhattacharya}}]{MarleyPRL74}
\bibinfo{author}{\bibfnamefont{A.~C.} \bibnamefont{Marley}},
  \bibinfo{author}{\bibfnamefont{M.~J.} \bibnamefont{Higgins}},
  \bibnamefont{and}
  \bibinfo{author}{\bibfnamefont{S.}~\bibnamefont{Bhattacharya}},
  \bibinfo{journal}{Phys. Rev. Lett.} \textbf{\bibinfo{volume}{74}},
  \bibinfo{pages}{3029} (\bibinfo{year}{1995}).

\bibitem[{\citenamefont{Olson and Reichhardt}(2000)}]{OlsonPRB61}
\bibinfo{author}{\bibfnamefont{C.~J.} \bibnamefont{Olson}} \bibnamefont{and}
  \bibinfo{author}{\bibfnamefont{C.}~\bibnamefont{Reichhardt}},
  \bibinfo{journal}{Phys. Rev. B} \textbf{\bibinfo{volume}{61}},
  \bibinfo{pages}{R3811} (\bibinfo{year}{2000}).

\bibitem[{\citenamefont{Reichhardt and Olson}(2002)}]{ReichhardtPRB65}
\bibinfo{author}{\bibfnamefont{C.}~\bibnamefont{Reichhardt}} \bibnamefont{and}
  \bibinfo{author}{\bibfnamefont{C.~J.} \bibnamefont{Olson}},
  \bibinfo{journal}{Phys. Rev. B} \textbf{\bibinfo{volume}{65}},
  \bibinfo{pages}{094301} (\bibinfo{year}{2002}).

\bibitem[{\citenamefont{Markovi\ifmmode~\acute{c}\else \'{c}\fi{}
  et~al.}(2000)\citenamefont{Markovi\ifmmode~\acute{c}\else \'{c}\fi{}, Dohmen,
  and van~der Zant}}]{Markovic}
\bibinfo{author}{\bibfnamefont{N.}~\bibnamefont{Markovi\ifmmode~\acute{c}\else
  \'{c}\fi{}}}, \bibinfo{author}{\bibfnamefont{M.~A.~H.} \bibnamefont{Dohmen}},
  \bibnamefont{and} \bibinfo{author}{\bibfnamefont{H.~S.~J.}
  \bibnamefont{van~der Zant}}, \bibinfo{journal}{Phys. Rev. Lett.}
  \textbf{\bibinfo{volume}{84}}, \bibinfo{pages}{534} (\bibinfo{year}{2000}).

\bibitem[{\citenamefont{Perruchot et~al.}(2000)\citenamefont{Perruchot,
  Williams, Mellor, Gaal, Sas, and Henini}}]{Perruchot}
\bibinfo{author}{\bibfnamefont{F.}~\bibnamefont{Perruchot}},
  \bibinfo{author}{\bibfnamefont{F.~I.~B.} \bibnamefont{Williams}},
  \bibinfo{author}{\bibfnamefont{C.~J.} \bibnamefont{Mellor}},
  \bibinfo{author}{\bibfnamefont{R.}~\bibnamefont{Gaal}},
  \bibinfo{author}{\bibfnamefont{B.}~\bibnamefont{Sas}}, \bibnamefont{and}
  \bibinfo{author}{\bibfnamefont{M.}~\bibnamefont{Henini}},
  \bibinfo{journal}{Physica B} \textbf{\bibinfo{volume}{284-288}},
  \bibinfo{pages}{1984} (\bibinfo{year}{2000}).

\bibitem[{\citenamefont{Altounian et~al.}(1994)\citenamefont{Altounian, Dantu,
  and Dikeakos}}]{AltounianPRB49}
\bibinfo{author}{\bibfnamefont{Z.}~\bibnamefont{Altounian}},
  \bibinfo{author}{\bibfnamefont{S.}~\bibnamefont{Dantu}}, \bibnamefont{and}
  \bibinfo{author}{\bibfnamefont{M.}~\bibnamefont{Dikeakos}},
  \bibinfo{journal}{Phys. Rev. B} \textbf{\bibinfo{volume}{49}},
  \bibinfo{pages}{8621} (\bibinfo{year}{1994}).

\bibitem[{\citenamefont{Hilke et~al.}(2003)\citenamefont{Hilke, Reid, Gagnon,
  and Altounian}}]{HilkePRL91}
\bibinfo{author}{\bibfnamefont{M.}~\bibnamefont{Hilke}},
  \bibinfo{author}{\bibfnamefont{S.}~\bibnamefont{Reid}},
  \bibinfo{author}{\bibfnamefont{R.}~\bibnamefont{Gagnon}}, \bibnamefont{and}
  \bibinfo{author}{\bibfnamefont{Z.}~\bibnamefont{Altounian}},
  \bibinfo{journal}{Phys. Rev. Lett.} \textbf{\bibinfo{volume}{91}},
  \bibinfo{pages}{127004} (\bibinfo{year}{2003}).

\bibitem[{\citenamefont{Kes and Tsuei}(1983)}]{KesPRB28}
\bibinfo{author}{\bibfnamefont{P.~H.} \bibnamefont{Kes}} \bibnamefont{and}
  \bibinfo{author}{\bibfnamefont{C.~C.} \bibnamefont{Tsuei}},
  \bibinfo{journal}{Phys. Rev. B} \textbf{\bibinfo{volume}{28}},
  \bibinfo{pages}{5126} (\bibinfo{year}{1983}).

\bibitem[{\citenamefont{Lefebvre et~al.}(2006)\citenamefont{Lefebvre, Hilke,
  Gagnon, and Altounian}}]{LefebvrePRB74}
\bibinfo{author}{\bibfnamefont{J.}~\bibnamefont{Lefebvre}},
  \bibinfo{author}{\bibfnamefont{M.}~\bibnamefont{Hilke}},
  \bibinfo{author}{\bibfnamefont{R.}~\bibnamefont{Gagnon}}, \bibnamefont{and}
  \bibinfo{author}{\bibfnamefont{Z.}~\bibnamefont{Altounian}},
  \bibinfo{journal}{Phys. Rev. B} \textbf{\bibinfo{volume}{74}},
  \bibinfo{eid}{174509} (pages~\bibinfo{numpages}{4}) (\bibinfo{year}{2006}),
  \urlprefix\url{http://link.aps.org/abstract/PRB/v74/e174509}.

\bibitem[{\citenamefont{Feigel\char39{}man
  et~al.}(1989)\citenamefont{Feigel\char39{}man, Geshkenbein, Larkin, and
  Vinokur}}]{FeigelmanPRL63}
\bibinfo{author}{\bibfnamefont{M.~V.} \bibnamefont{Feigel\char39{}man}},
  \bibinfo{author}{\bibfnamefont{V.~B.} \bibnamefont{Geshkenbein}},
  \bibinfo{author}{\bibfnamefont{A.~I.} \bibnamefont{Larkin}},
  \bibnamefont{and} \bibinfo{author}{\bibfnamefont{V.~M.}
  \bibnamefont{Vinokur}}, \bibinfo{journal}{Phys. Rev. Lett.}
  \textbf{\bibinfo{volume}{63}}, \bibinfo{pages}{2303} (\bibinfo{year}{1989}).

\bibitem[{\citenamefont{Fisher}(1989)}]{FisherPRL62}
\bibinfo{author}{\bibfnamefont{M.~P.~A.} \bibnamefont{Fisher}},
  \bibinfo{journal}{Phys. Rev. Lett.} \textbf{\bibinfo{volume}{62}},
  \bibinfo{pages}{1415} (\bibinfo{year}{1989}).

\end{thebibliography}

\end{document}